\documentclass[aps,twocolumn,nofootinbib,showpacs]{revtex4}
\usepackage{pifont}

\usepackage{amsmath}
\usepackage{amsfonts}
\usepackage{amssymb}
\usepackage{color}
\usepackage{graphicx}
\usepackage{mathrsfs}
\usepackage{ifpdf}

\ifpdf   
  \RequirePackage[pdftex]{hyperref}
\else    
    \RequirePackage[dvips]{hyperref}
  \fi
\hypersetup{bookmarksnumbered,%
              colorlinks,%
               linkcolor=blue,%
               citecolor=blue,%
              plainpages=false,%
            pdfstartview=FitH}

\def\bc{\begin{center}}
\def\nno{\nonumber}
\def\ec{\end{center}}
\def\be{\begin{eqnarray}}
\def\ee{\end{eqnarray}}

\def\dS{dS}
\definecolor{dyellow}{rgb}{1.,0.8,.0}
\definecolor{myblue}{rgb}{.1,.1,.7}
\definecolor{dcyan}{rgb}{.0,.6,.6}
\definecolor{dmagenta}{rgb}{0.6,0.0,0.6}
\definecolor{brown}{rgb}{0.6,0.2,0.}
\definecolor{darkblue}{rgb}{.0,.0,0.5}
\definecolor{darkred}{rgb}{0.75,0.0,0.0}
\definecolor{orange}{rgb}{1.,.6,.0}
\definecolor{dorange}{rgb}{0.8,.4,.0}
\definecolor{darkgreen}{rgb}{0.0,0.6,0.0}
\definecolor{purple}{rgb}{.4,.0,.4}
\definecolor{lightgrey}{rgb}{0.7, 0.7, 0.7}
\definecolor{grey}{rgb}{0.4, 0.4, 0.4}
\def\black{\color{black}}

\def\al{\alpha}
\def\ga{\gamma}
\def\dl{\delta}
\def\eps{\epsilon}
\def\ka{\kappa}
\def\la{\lambda}
\def\th{\theta}
\def\si{\sigma}

\def\La{\Lambda}

\def\d#1#2{\frac{\displaystyle #1}{\displaystyle #2}}

\newcommand\btd{\raise 2pt
\hbox{$\hat\bigtriangledown$}\hskip 1.5pt}
\newcommand\bt{\raise 2pt
\hbox{$\bigtriangledown$}\hskip 1.5pt}

\newcommand{\omits}[1]{}
\newcommand{\cR}{{\cal R}}

\def\PRD{{Phys. Rev.}~{\bf D}}
\def\PRL{{Phys. Rev. Lett. }}
\def\PLA{{Phys. Lett.}~{\bf A}}

\begin{document}

\title{On Torsion-free Vacuum Solutions of the Model of de Sitter Gauge Theory of Gravity (II)}

\author{Chao-Guang Huang$^{a,b}$\footnote{Email: huangcg@ihep.ac.cn} and
Meng-Sen Ma$^{a,c}$\footnote{Email: mams@ihep.ac.cn}}

\medskip

\affiliation{\footnotesize $^a$ Institute of High Energy Physics, Chinese
Academy of Sciences,
Beijing 100049, China \\
$^b$ Theoretical Physics Center for Science Facilities, Chinese Academy of
Sciences, Beijing 100049, China \\
$^c$ Graduate School of Chinese Academy of
Sciences, Beijing, 100049, China}

\begin{abstract}
It is shown that all torsion-free vacuum solutions of the model of dS
gauge theory of gravity are the vacuum solutions of Einstein
field equations with the same positive cosmological constant.
Furthermore, for the gravitational theories with more general quadratic
gravitational Lagrangian ($F^2+T^2$), the torsion-free vacuum solutions are
also the vacuum solutions of Einstein field equations.

\end{abstract}


\maketitle

\bigskip



The astronomical observations show that our universe is probably an
asymptotically de Sitter(dS) one \cite{SN,WMAP}. It raises the
interests on dS gauge theories of gravity. There is a model of dS
gravity{\footnote{Hereafter, the model of dS gauge theory of gravity
is called the dS gravity for short in this paper.}}, which was first
proposed in the 1970's \cite{dSG, T77}.  The dS gravity can be
stimulated from dS invariant special relativity \cite{dSSR,
meetings, dSSR2} and the principle of localization --- the full
symmetry of the special relativity as well as the laws of dynamics
are both localized \cite{Guo2,vacuum,cosmos} --- and realized in
terms of the dS connection on a kind of totally umbilical
submanifolds (under the dS-Lorentz gauge) and Yang-Mills type action
\cite{dSG,Guo2,cosmos}.  It has been shown \cite{cosmos} that the dS
gravity may explain the accelerating expansion and supply a natural
transit from decelerating expansion to accelerating expansion
without the help of the introduction of matter fields in addition to
dust. The different de Sitter spacetimes with nonzero torsion in the
dS gravity have been presented in \cite{HMZ}.  It has also been
shown that all vacuum solutions of Einstein field equations with a
cosmological constant are the vacuum solutions of the set of field
equations without torsion \cite{Guo2, vacuum}. In particular,
Schwarzschild-dS and Kerr-dS metrics are two solutions.  The purpose
of the present letter is to show that all vacuum, torsion-free
solutions in the dS gravity are the vacuum solutions of Einstein
field equations with the same positive cosmological constant.
Therefore, one may expect that the dS gravity may pass all
solar-system-scale observations and experimental tests for general
relativity (GR).

The dS gauge theory of gravity is established based on the following
consideration.  The non-gravitational theory is de Sitter invariant special
relativity.  The theory of gravity should follow the principle of
localization, which says that the {\it full symmetry} as well as the {\it laws
of dynamics} are both localized, and the gravitational action takes
Yang-Mills-type.

A model of dS gauge theory of gravity has been constructed
\cite{dSG,T77,Guo2,vacuum,cosmos} in terms of the de Sitter connection in the
\dS-Lorentz frame, which reads\footnote{The same connection with different
gravitational dynamics has also been studied (See, e.g.
\cite{MM,SW,Wil,FS,AN,Lec,Wise,Mahato,tresguerres})}
\be\label{dSc}%
({\cal B}^{AB}_{\ \ \ \,{\mu}})=\left(
\begin{array}{cc}
B^{ab}_{\ \ \,{\mu}} & R^{-1} e^a_\mu \smallskip\\
-R^{-1}e^b_\mu &0
\end{array}
\right ) \in \mathfrak{so}(1,4).
\ee%
where ${\cal B}^{AB}_{\ \ \ \, \mu}=\eta^{BC}{\cal B}^A_{\ \, C\mu}$
.  Its curvature is then
\be\label{dSLF}%
 ({\cal F}^{AB}_{~~~\mu\nu})%
=\left(
\begin{array}{cc}
F^{ab}_{~~\mu\nu} + R^{-2}e^{ab}_{~~ \mu\nu} & R^{-1} T^a_{~\mu\nu}\\
-R^{-1}T^b_{~\mu\nu} &0
\end{array}
\right ), 
\ee%
where $e^a_{~b\mu\nu}=e^a_\mu e_{b\nu}-e^a_\nu e_{b\mu},
e_{a\mu}=\eta_{ab}e^b_\mu$, $ F^{ab}_{~~ \mu\nu}$ and $
T^a_{~\mu\nu}$ are the curvature and torsion of the Lorentz
connection which are defined as:
\be%
&&\begin{split}
&\Omega^a=d\th^a+\theta^a_{~b}
\wedge\th^b=\frac{1}{2}T^a_{~\mu\nu}dx^\mu\wedge
dx^\nu  \\%
&T^a_{~\mu\nu}=\partial_\mu
e^a_\nu-\partial_\nu e^a_ \mu+B^a_{~c \mu}e^c_\nu-B^a_{~c
\nu}e^c_\mu,
\end{split}\\%
&&\begin{split}
&\Omega^a_{~b}=d\theta^a_{~b}+\theta^a_{~c}\wedge\theta^c_{~b}
=\frac{1}{2}F^a_{~b
\mu\nu}dx^\mu\wedge dx^\nu ,\\ %
&F^a_{~b
\mu\nu}=\partial_\mu B^a_{~b\nu} -\partial_ \nu
B^a_{~b\mu}+B^a_{~c\mu}B^c_{~b
\nu}-B^a_{~c\nu}B^c_{~b\mu},   %
\end{split}
\ee%
where
$\theta^a=e^a_{~\mu}dx^{\mu},\theta^a_{~b}=B^a_{~b\mu}dx^{\mu}$.

The action for the model of de Sitter gauge theory of gravity with
sources takes
the form of%
\be\label{S_t}%
S_{\rm T}=S_{\rm GYM}&+&S_{\mathbf M},
\ee%
where
\be\label{GYM}%
S_{\rm GYM}&=&\frac{\hbar}{4g^2}\int_{\cal M}d^4 x e {\bf Tr}_{dS}({\cal
F}_{\mu\nu}{\cal F}^{\mu\nu})\ \ \
\nno \\
&=& -\int_{\cal M}d^4x e
\left[{\frac{\hbar}{4{\black g^2}} F^{ab}_{~\mu\nu}F_{ab}^{~\mu\nu}}
-{\chi(F-2\Lambda)} \right . \nno \\
&&\left . \qquad \qquad - \frac{\chi}{2} T^a_{~\mu\nu}T_a^{~\mu\nu}\right] 
\ee%
is the gravitational Yang-Mills action and $S_M$ is the action of sources with
minimum coupling.  In Eq.(\ref{GYM}), $g=\hbar^{1/2}R^{-1}\chi^{-1/2}\sim 10^{-61}$ is
the dimensionless gravitational coupling constant, $e=\det(e^a_\mu)$, $\La =
3/R^2$, $\chi=1/({ 16}\pi G)$, $G$ is the
Newtonian gravitational coupling constant, $F= -\frac{1}{2} F^{ab}_{\ \
\mu\nu}e_{ab}^{\ \ \mu\nu}$ is the scalar curvature of the Cartan connection.
($c=1$.)

The field equations can be given via the variational principle with respect to
$e^a_{~\mu},B^{ab}_{~~\mu}$,
\be%
&&T_{a~~\ ||\nu}^{~\mu\nu } - F_{~a}^\mu+\frac{1}{2}F e_a^\mu - \Lambda
e_a^\mu = 8\pi G( T_{{\rm M}a}^{~~\mu}+T_{{\rm G}a }^{~~\mu}),\nno \\
&&\label{Geq2}\\
&&F_{ab~~\ ||\nu}^{~~\mu\nu} = R^{-2}(16\pi G S^{\quad \mu}_{{\rm M}ab}
+S^{\quad \mu}_{{\rm G}ab}).\label{Geq2'}
\ee
Here, $||$ represents the covariant derivative defined by Christoffel symbol
$\{^\mu_{\nu\ka}\}$ and Lorentz connection $B^a_{\ b\mu}$,
$F_a^{~\mu}=-F_{ab}^{~~\mu\nu}e^b_\nu$. 
\be T_{{\rm M}a}^{~~\mu}=-\d 1 e \d {\dl S_{\rm M}}{\dl e^a_\mu}, \qquad
S_{{\rm M}ab}^{\quad \, \mu}=\d 1 {2\sqrt{-g}}\d {\dl S_{\rm M} }{\dl
B^{ab}_{\ \ \mu}}  \ee
are the tetrad form of the stress-energy tensor and spin current for matter
field, respectively.
\be
T_{{\rm G}a}^{~~\mu} = \hbar g^{-2} T_{{\rm F}a}^{~~\mu}+2\chi T_{{\rm
T}a}^{~~\mu}  %
\ee
is the tetrad form of the stress-energy tensor of gravitational
field, which can be split into the curvature part
\be
\label{emF}
T_{{\rm F}a}^{~~\mu}=\omits{-\frac{1}{4e} \frac{\delta} {\delta e^a_{\mu}}\int
d^4x e {\rm Tr}(F_{\nu\ka}F^{\nu\ka}) =} e_{a}^\ka {\rm Tr}(F^{\mu \la}F_{\ka
\la})-\frac{1}{4}e_a^\mu {\rm Tr}(F^{\la \si} F_{\la \si})  \ee and torsion
part
\be%
T_{{\rm T}a}^{~~\mu}=\omits{-\frac{1}{4e} \frac{\delta} {\delta e^a_{\mu}}
\int d^4x e T^b_{\ \nu\ka}T_b^{\ \nu\ka} =} e_a^\ka
T_b^{~\mu\la}T^{b}_{~\ka\la}-\frac{1}{4}e_a^\mu T_b^{~\la\si}T^b_{~\la\si}.
\ee%
Similarly, the gravitational spin-current
\be\label{spG}%
 S_{{\rm G}ab}^{\quad \, \mu}=S_{{\rm F}ab}^{\quad \,
\mu}+2S_{{\rm
T}ab}^{\quad \,\mu} %
\ee%
can also be divided into two parts
\be%
S_{{\rm F}ab}^{\quad \, \mu}&=&{-}e^{~~\mu \nu}_{ab\ \
{||}\nu} = Y^\mu_{~\, \la\nu
e_{ab}^{~~\la\nu}}+Y ^\nu_{~\, \la\nu } e_{ab}^{~~\mu\la},  \\
S_{{\rm T}ab}^{\quad \mu}&=& T_{[a}^{~\mu\la}e_{b]\la}^{},
\ee%
where
\be Y^\la _{~~\mu\nu}= \d 1 2 (T^\la _{\ \,\nu\mu}+T^{\ \la} _{\mu \
\,\nu} +T^{\ \la} _{\nu \ \,\mu}) \ee
is the contortion.

For vacuum, torsion-free cases, ${\cal F}_{ab}^{~~\mu\nu}$ reduces to
the Riemann curvature ${\cal R}_{ab}^{~~\mu\nu}$ and  the
gravitational field equations reduce to
\be
 && {\cal R}_{~a}^\mu-\frac{1}{2}{\cal R} e_a^\mu +
\Lambda
e_a^\mu = -8\pi G T_{{\rm R}a }^{~~\mu}, \label{ElEq}\\%
&&{\cal R}_{ab~~;\nu}^{~~\mu\nu} = 0, \label{YangEq} \ee
where $T_{{\rm R}a}^{~~ \mu}=e_{a}^{\nu}T_{{\rm R}\ \nu}^{~\mu}$
is the torsion-free case of $T_{{\rm F}a}^{~~ \mu}$ and called as the
tetrad form of the stress-energy tensor of Riemann curvature ${\cal
R}_{ab}^{~~\mu\nu}$, and a semicolon $;$ is the covariant derivative
defined by the Christoffel symbols and Ricci rotation coefficients.
 Eq.
(\ref{ElEq}) is the Einstein-like equation, while Eq.(\ref{YangEq})
is the Yang equation in Stephenson-Kilmister-Yang theory of gravity
\cite{Yang}.

The trace of Eq.(\ref{ElEq}) gives%
\be%
\cR=4\La. %
\ee%
It can be shown \cite{WZC,vacuum}  that
\be%
T_{{\rm R} \mu}^{~~\nu} =2C_{\la\mu}^{~~\ka\nu}{\cal S}^\la_\ka +\frac{{\cal
R}}{3}{\cal S}_\mu^\nu , \label{emR}
\ee%
where $C_{\la\mu\ka\nu}$ is the Weyl tensor, %
\be %
{\cal S}_{\mu\la}={\cal R}_{\mu\nu}-\frac 1 4 {\cal R}g_{\mu\nu} %
\ee%
is the traceless Ricci tensor.  Therefore, Eq.(\ref{ElEq}) after multiplied by
$e^a_\nu$ can be rewritten as %
\be\label{main} %
(\chi + \frac 2 3 \La ){\cal S}_{\mu\nu}+
C_{\mu\ka\nu\la}{\cal S}^{\ka\la}=0. %
\ee%

Since the torsionless curvature tensor is symmetric with respect to its
two-pair indices and satisfies the Bianchi identity, Eq.(\ref{YangEq})
multiplied by $e^a_\ka e^b_\la$ is equivalent to
\be%
{\cal R}^\mu_{\ \ka;\la}={\cal R}^\mu_{\ \la;\ka}. \nno
\ee%
For the constant Ricci scalar, it can be rewritten as
\be%
{\cal S}^\mu_{\ \ka;\la}={\cal S}^\mu_{\ \la;\ka}. \label{yang}
\ee%
Bianchi identity together with the irreducible decomposition of Riemann curvature
tensor and Eq.(\ref{yang}) results in%
\be \label{WeylB} %
C_{\mu\nu[\la\si; \ka]}=0. %
\ee%
From Eq.(\ref{yang}), Ricci identity, and the irreducible decomposition of
Riemann curvature tensor, one can obtain%
\be%
0&=&({\cal S}^\mu_{\ \ka;\la \rho}-{\cal S}^\mu_{\ \la;\ka \rho}) + ({\cal
S}^\mu_{\ \rho;\ka\la }-{\cal S}^\mu_{\ \ka;\rho\la})\nno\\
&&+({\cal S}^\mu_{\ \la
;\rho\ka } -{\cal S}^\mu_{\ \rho; \la\ka })\nno \\
&=&{C}^\mu_{\ \si\la\rho }{\cal S}^\si_{\ \ka}  +{C}^\mu_{\
\si\ka\la }{\cal S}^\si_{\ \rho}  +{C}^\mu_{\ \si\rho\ka }{\cal S}^\si_{\
\la}.\nno
\ee%
This is the integrability condition of Eq.(\ref{YangEq}) %
\be \label{IC}%
C^*_{\ \mu\lambda\nu\ka}{\cal S}^{\la\ka}=0,  %
\ee
where $^*$ is the Hodge star.

Eqs.(\ref{main}), (\ref{WeylB}), (\ref{IC}) and the traceless condition of
${\cal S}_{\mu\nu}$ constitute a set of equations for vacuum, torsion-free
solutions of the dS gravity.  The set of the equations are the same as those
in Ref.\cite{DFS} except $\cR_{\mu\nu}$'s have been replaced by ${\cal
S}_{\mu\nu}$'s.

The traceless Ricci tensor ${\cal S}_{\mu\nu}$ can be classified into 4
algebraic general types in terms of their eigenvalues \cite{exact}.  In
Segr\'e notation, they are $[1,1\,1\,1]$, $[Z \,\bar Z,1\,1]$, $[2,1\,1]$, and
$[3,1]$. In the orthogonal tetrad, the traceless Ricci tensors are given as
follows. For [1,1\,1\,1] type,
\be%
{\cal S}_{ab}=  \left(\begin{array}{cccc}
                               a &  &  &  \\
                                & -b &  &  \\
                                &  & -c &  \\
                                &  &  & -d \\
                             \end{array}\right)
\ee
with $a+b+c+d=0$.  For [$Z \,\bar Z$,1\,1] type,
\be
{\cal S}_{ab}&=&
\left(
         \begin{array}{cccc}
           f & -g &  &  \\
           -g & -f &  &  \\
            &  & -c &  \\
            &  &  & -d \\
         \end{array}
       \right)
\ee
with $2f+c+d=0$.  For\ [2,1\,1]\ type,
\be %
{\cal S}_{ab}&=& \left(
                       \begin{array}{cccc}
                         \pm 1+a & \pm 1 &  &  \\
                         \pm 1 & \pm 1-a &  &  \\
                          &  & -c &  \\
                          &  &  & -d \\
                       \end{array}
                     \right)
\ee
with $2a+c+d=0$.  For [3,1]\ type,
\be
{\cal S}_{ab}&=& \left(
                      \begin{array}{cccc}
                        a &  & 1 &  \\
                         & -a & 1 &  \\
                        1 & 1 & -a &  \\
                         &  &  & -d \\
                      \end{array}
                    \right)
\ee
with $3a+d=0$.

In the null tetrad $\pmb{l}=\frac {\sqrt{2}} 2(\pmb{e}^0+\pmb{e}^1)$,
$\pmb{n}=\frac {\sqrt{2}} 2(\pmb{e}^0-\pmb{e}^1)$, $\pmb{m}=\frac {\sqrt{2}}
2(\pmb{e}^2+i\pmb{e}^3)$, they are
\be%
{\cal S}_{a'b'}=2\left(\begin{array}{cccc}
                               \Phi_{00} & \Phi_{11} &  &  \\
                               \Phi_{11} & e\Phi_{00}&  &  \\
                                &  & \Phi_{02} & \Phi_{11} \\
                                &  & \Phi_{11} & \Phi_{02} \\
                             \end{array}
                           \right)
\ee%
with $e =+1$, $\Phi_{00} = a - b$, $\Phi_{02}= d-c $, $\Phi_{11}= a + b = -c
-d$ for $[1,1\,1\,1]$ type and with $e = - 1$, $\Phi_{00}= 2g$, $\Phi_{02} =
d-c $, $\Phi_{11} = 2f = -c - d$ for $[Z \,\bar Z,1\,1]$ type;
\be%
{\cal S}_{a'b'}=2\left(
                       \begin{array}{cccc}
                         \pm 1 & \Phi_{11} &  &  \\
                         \Phi_{11} & 0 &  &  \\
                          &  & \Phi_{02} & \Phi_{11} \\
                          &  & \Phi_{11} & \Phi_{02} \\
                       \end{array}
                     \right)%
\ee%
with $\Phi_{11}=2a=-c-d=0$ and $\Phi_{02}=d-c$ for $[2,1\,1]$ type; and%
\be%
{\cal S}_{a'b'}=2\left(
                      \begin{array}{cccc}
                        0 & \Phi_{11} & \frac 1 2 & \frac 1 2 \\
                        \Phi_{11} 0 & 0 & 0 \\
                        \frac 1 2 & 0 & -2\Phi_{11} &  \Phi_{11} \\
                        \frac 1 2 & 0 & \Phi_{11} & -2\Phi_{11} \\
                      \end{array}
                    \right)%
\ee%
with $\Phi_{11}=2a=-a-d=0=(a-d)/2$ for $[3,1]$ type.

Following the analysis of Debney et al in \cite{DFS}, one
can prove%
\be%
{\cal S}_{\mu\nu}=0. %
\ee%
This is nothing but the vacuum Einstein field equation with the
cosmological constant $\La$.

In conclusion, all vacuum, torsion-free solutions in the dS gravity
must be the solutions of vacuum Einstein field equation with the same
positive cosmological constant.  Together with the conclusion
obtained in \cite{vacuum}, we conclude that the set of vacuum,
torsion-free solutions in the dS gravity and the solutions of vacuum
Einstein field equations with the same cosmological constant are
equivalent.  Therefore, the dS gravity is expected to pass the observational
tests on the scale of a solar system and explain the indirect
evidence of the existence of gravitational wave from the observation
data on the binary pulsar PSR1913+16.

Furthermore, the conclusion can be generalized to the Lagrangian
\be%
S_G&=& \int d^4x\{\chi (F-2\La)+\al F_{ab}^{\ \ \,\mu\nu}F^{ab}_{\ \ \,\mu\nu}+\beta T^a_{\
\mu\nu}T_a^{\ \mu\nu} \nno \\
&& + \eps e^{ab}_{\ \ \, \la\si}e^{cd}_{\ \ \, \mu\nu}F_{ab}^{\ \
\,\mu\nu}F_{cd}^{\ \ \,\la\si}
+\ka e^b_{\si} e_{c}^{\mu}F_{ab\mu\nu}F^{ac \,\nu\si}\nno\bigskip \\
&&+ \ga F_a^{\ \,\mu}F^a_{\ \,\mu}
+\dl e^a_\nu e^b_\mu F_a^{\ \,\mu}F_b^{\ \,\nu}
+\la e_a^\la e_b^{\si} T^a_{\ \mu\la}T^{b\mu}_{\ \ \si}\nno \\
&&
+\si e_a^\si e_b^{\mu} T^a_{\ \mu\la}T^{b\la}_{\ \ \si}
\}. \nno %
\ee%
\omits{unless
\be \label{exception}
\al+4\eps-\frac 1 2 \ka=0.
\ee}
The reasons are as follows.  For the torsion-free case, the last two terms have no contribution to the
vacuum, torsion-free field equations, and the two terms in the second line
contribute the
same term in the field equations as $F _{ab}^{\ \ \mu\nu} F^{ab}_{\ \ \mu\nu}$
does, thus only alter the unimportant coefficients.  The above argument is
not valid only when the coupling constants $\al$, $\eps$, and $\ka$ are suitably arranged
so that Eq.(\ref{YangEq}) does not appear.
The first two terms in the third line add the
term $(\cR^\mu_{[a}e^\nu_{b]})_{;\nu}$ in Yang equation and the stress-energy
tensor $\cR_{\mu\la}\cR^{\nu\la}-\frac 1 4\dl^\nu_\mu
\cR_{\si\la}\cR^{\si\la}$ in Einstein-like equation. The latter will not
change the constancy of Ricci scalar, and thus the former has still the form
of Eq.(\ref{yang}). In other words, Eqs. (\ref{WeylB}), (\ref{IC}) and the
traceless condition of ${\cal S}_{\mu\nu}$ are not changed.  Only
Eq.(\ref{main}) has slightly different form.   A detailed calculation shows
that it will not destroy the conclusion.

Obviously, the conclusion is still valid if the integral of the
second Chern form of the dS connection over the manifold is added in
the action. Finally, the similar discussions can be applied to the
AdS case as well and the conclusion is still valid if they can be
generalized to the AdS gravity.

\begin{acknowledgments}\vskip -4mm
This work is supported by NSFC under Grant Nos. 10775140 and Knowledge
Innovation Funds of CAS (KJCX3-SYW-S03).
\end{acknowledgments}

\end{document}